\documentclass{aa}  
\usepackage{graphicx}
\usepackage{txfonts}
\def\integral{{\it INTEGRAL}}
\def\chandra{{\it Chandra}}

\def\sax{{\it BeppoSAX}}
\def\xmm{{\it XMM-Newton}}

\def\1e{{\mbox{1E~1743.1--2843}}}

\newcommand{\g}{$\gamma$}

\begin{document}
   \title{First broad band study of the mysterious source 1E 1743.1--2843\thanks{Based on observations with \integral, an ESA project with instruments and science data centre funded by ESA member states (especially the PI countries: Denmark, France, Germany, Italy, Switzerland, Spain), Czech Republic and Poland, and with participation of Russia and the USA.}}


   \author{M. Del Santo \inst{1}, L. Sidoli \inst{2}, A. Bazzano \inst{1}, M. Cocchi \inst{1}, G. De Cesare \inst{1,3,4},
A. Paizis \inst{2}, P. Ubertini \inst{1}} 

   \offprints{melania.delsanto@iasf-roma.inaf.it}

\institute{ Istituto di Astrofisica Spaziale e Fisica Cosmica di Roma -- INAF, via del Fosso del Cavaliere 100, 40133 Roma, Italy
\and Istituto di Astrofisica Spaziale e Fisica Cosmica di Milano -- INAF, via E. Bassini 15, 20133 Milano, Italy
\and Dipartimento di Astronomia, Universit\'a degli Studi di Bologna, via Ranzani 1, 40127 Bologna, Italy
\and Centre d'\'Etude Spatiale des Rayonnements, CNRS/UPS, B.P. 4346, 31028 Toulouse Cedex 4, France 
}

\authorrunning{M. Del Santo et al.}
\titlerunning{First high energy study of the mysterious source 1E 1743.1--2843}

   \date{Received...; accepted...}

 
  \abstract
   {In the last years, the persistent source \1e has been observed in the X-rays, but never above 20 keV. 
In previous works, it was stressed that a possible high energy emission could give further indications 
on the accreting object nature which remains still unknown.}
{We present here more than two years of \1e monitoring with \integral/IBIS as well as
public \xmm~ and \chandra~ X-ray observations.}
   {The temporal study in the 20--40 keV band shows a rather constant flux on few months time scale.
Based on this result we have performed the broad-band spectral analysis using EPIC/IBIS non simultaneous data 
and ACIS-I/IBIS data collected during 2004.}
   {In $\sim$2 Ms, we report a detection of 6$\sigma$ in the energy range 35--70 keV. 
The first broad-band study (2--70 keV) shows a steep slope ($\sim$3) and a black body temperature of 1.7 keV.}
   {Combining spectral parameters and discussion about the luminosity evaluations for different possible distances,
our conclusions are in favour of a LMXB system with a neutron star at distance higher than the Galactic Centre,
even though a firm conclusion can not be stated.} 


   \maketitle

\section{Introduction}
Among the numerous X-ray binaries populating the Galactic Centre (most of them
transients and highly variable), the source \1e 
presents a peculiar behaviour because of its X-ray persistent flux and the lack of strong variability. 
The measured $N_{H}$ value larger than $10^{23}$ cm$^{-2}$ suggested a distance similar to 
the GC, or even greater (\cite{cremonesi99}).
Until now, in spite of the long \sax~monitoring programme of the GC region,
no bursting activity has ever been reported with the Wide Field Cameras for the \1e source (\cite{zand04}).
Cremonesi et al. (1999) ruled out the neutron star HMXB nature because of the absence of pulsations and/or eclipses.                          .
\xmm~observation reported by Porquet et al. (2003) has provided the best position
of the source and proposed that \1e might be explained in terms of a black hole candidate in low/hard state,
underlying that high energy observations could help in the determination of the compact object nature.  

Due to the unprecedented combined spatial resolution and sensitivity of the \g-ray imager IBIS (\cite{ubertini03})
on-board \integral~satellite (\cite{winkler03}),
the hard X-ray detection (20--40 keV) of \1e has been reported for the first time in the 
IBIS/ISGRI soft \g-ray catalogue at $\sim$5 mCrab flux level (\cite{bird04}).
In this work, we present the first \1e broad-band study up to 70 keV, 
obtained during more than two years of monitoring programme with IBIS  
on-board the \integral~satellite. Moreover, we present a re-analysis of \xmm~public data  
and results of \chandra~public data never shown elsewhere.

\begin{table}[t!]
\caption{\integral, \xmm~and \chandra~observations log. Exposure times of MOS1 and MOS2 are reported for 
EPIC.} \label{tab:obs_log}
\begin{center}
\begin{tabular}[c]{llcc}
\hline\noalign{\smallskip}
Instrument & Orbits$^\dagger$&  Obs Time         & Exp\\
           &  /Obs$^\ddag$     &      &  (ks)        \\
\hline\noalign{\smallskip}
IBIS & 46-63  &      Feb-Apr 2003  &   $\sim$400       \\
               & 103-120  &      Aug-Oct 2003   &   $\sim$1400   \\
               & 164-185  &      Feb-Apr 2004  &   $\sim$900    \\
               & 229-249 &      Aug-Oct 2004   &   $\sim$160    \\
               & 291-307  &      Mar-Apr 2005   &   $\sim$44    \\
\noalign{\smallskip\hrule\smallskip}
EPIC  & 401     &  19 Sep 2000 &  29.1   \\
                &  501    &  21 Sep 2000 &  24.7   \\
\noalign{\smallskip\hrule\smallskip}
ACIS-I  & 2292    &  17 July 2001 &      11.6  \\
   & 2276    &  18 July 2001   &      11.6   \\
   & 4500    &  9 June 2004   &      98.6   \\
\noalign{\smallskip\hrule\smallskip}
\end{tabular}\\
\end{center}
\small{$^\dagger$ Orbits refer to \integral.}\\
\small{$^\ddag$ Observation number refers to \xmm~and \chandra.}
\end{table}

%

\section{Observations and data analysis}

\subsection{2000--2005 X-ray and soft \g-ray observations}
Since February 2003, \integral~has been observing the Galactic Centre.
We have analysed all the public data collected with the low energy detector layer, i. e. ISGRI (\cite{lebrun03}),
performed in 2003, 2004, and the first part of the 2005 (Tab. \ref{tab:obs_log}).
The small field of view of the X-ray monitor on-board \integral~(JEM-X) combined with
the observing strategy of the mission characterised by many off-axis pointings,
do not allow for useful JEM-X scientific pointings and so 
we searched for archival recent observations of the source field
performed with different X-ray missions.
This search resulted in two \xmm~observations performed in 2000 and  
three \chandra/ACIS-I observations performed in 2001 and 2004 (Tab. \ref{tab:obs_log}). 
Among these, only the results of one of the \xmm~observations 
have already been published (\cite{porquet03}).
The ACIS-I/\chandra observations had  
the Arches Cluster as main target, thus \1e~was always observed off-axis.

\subsection{IBIS, EPIC and ACIS-I scientific analysis}
The IBIS/ISGRI data have been analysed using the release 5.0 of the \integral~off-line analysis software (OSA).  
The 20--40 keV light curve has been extracted from mosaicked images of each GC visibility window. 
Two average spectra have been obtained, one for the 2003 and the second one for 2004. 
The count rate spectra have been extracted from the image mosaics obtained in three energy bands:
15--20.7 keV, 20.7--34.1 keV, 34.1--70.4 keV.

\xmm~data have been reprocessed with the version 6.5 of the Science Analysis
Software (SAS) and known hot, or flickering, pixels and electronic
noise were rejected. 
To reduce pile-up found in both MOS and PN data,
we extracted spectra excluding the core of the PSF.
An annulus with an internal radius of 10$''$ and an outer radius of 40$''$ has been chosen.
We verified with {\em epatplot} tool
that in the extracted events from these coronae, the pile-up effect was no more present.
Background spectra were obtained from source free regions
of the same observations.

The events files (level 2) processed by the \chandra~X-ray Centre and available from
the public archive, were used for the analysis.
The \chandra~Interactive Analysis of Observations 
(CIAO) tool version 3.2 was used. We checked that periods of anomalous background levels were absent.

Since \1e~is a relatively bright X-ray source 
(few 10$^{-10}$~erg~cm$^{-2}$~s$^{-1}$ in the 2--10~keV),
the observations suffered from severe pile-up effects of about 80\% for ACIS-I.
When we compared spectra extracted from the core
of the PSF with spectra taken from annular regions which excluded the central pixels,
a prominent hardening in the spectra extracted from the PSF core has been observed.
To avoid pile-up problems, we extracted counts from the readout streaks 
produced by photons collected during the $\sim$40~ms required to transfer
an image frame to the readout buffer.
We considered a boxed region of width $\sim$24$''$ centered on the streaks.

Spectral fitting and the estimation of fluxes used for the luminosities calculation
were done with XSPEC v. 11.3.2; single parameter 
uncertainties were calculated at the 90\% confidence level. 

   \begin{figure}[t!]
   \centering
   \includegraphics[width=4.1cm, angle=+90]{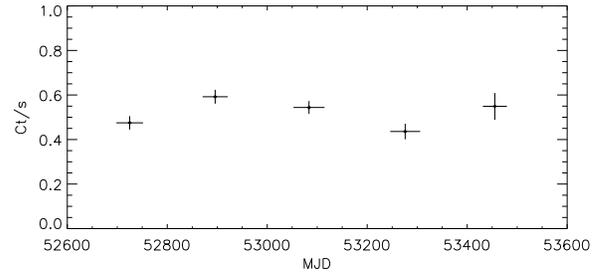}
      \caption{The 20--40 keV temporal behaviour shows marginal variability over few months 
time scale. The mean level of the flux is roughly 5 mCrab.}
         \label{fig:light}
   \end{figure}
%

\section{Results}
\subsection{X-ray spectra with \xmm~ and \chandra}
EPIC and ACIS-I background subtracted spectra have been fitted using different spectral models:
a simple power law, a single black body and a multi-color disc black body (\texttt{PO},
\texttt{BB} and \texttt{DISKBB} in XSPEC, respectively); in the three cases the hydrogen column density $N_{H}$
has been included as a free parameter (\texttt{PHABS} model).

An absorbed power law does not provide a good fit to the \xmm~data.
The best-fit is obtained with a single BB model, with a blackbody radius of 1.8$\pm$0.1~km 
and 2.0$\pm$0.1~km  and a temperature kT$_{bb}$
of $1.73\pm0.04$~keV and 1.56$\pm$0.04~keV, for the two XMM-Newton observations, respectively.

The fitting results with MCD confirm this trend, even though the 
multicolor disc temperatures (kT$_{diskbb}$)
are systematicly higher (2.9 $\pm~0.1$ and 2.4 $\pm~0.1$ keV 
for the two data set, respectively) than kT$_{bb}$.

Porquet et al. (2003) obtained a good reduced $\chi^2$ adding a simple power law to the thermal
component with a kT$_{bb} \sim$ 0.15 keV.
We tried this two-components model fitting obtaining a reduced $\chi^2$ = 1.3 (369), 
which is not significantly better than the value from the one-component model. 
This is due to the fact that the authors did not perform any correction for the pile-up (as they reported on).
In any case, a BB model with kT=0.1 keV 
would be completely hidden by the huge absorption. 

All spectral parameters of the observations performed in 2001 with ACIS-I
are consistent, within the errors, with the \xmm~data results.
Although compatible fluxes have been measured, a 
hardening in the X-ray spectrum
of 2004 seems to be present, with a 
higher temperature of kT$_{bb}$=$2.2 \pm 0.3 $ keV.         

   \begin{figure}[t]
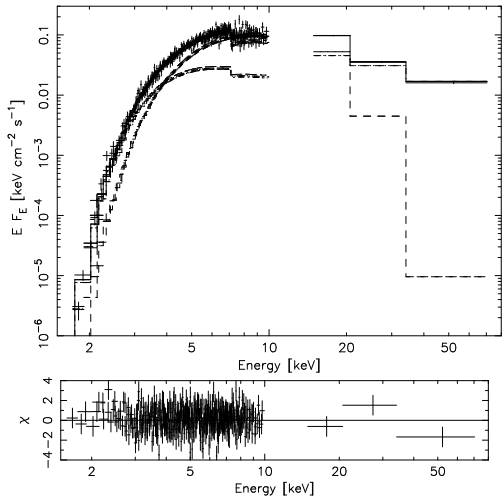

   \centering
   \includegraphics[width=4.9cm, angle=-90]{delsanto06_fig2.ps}
   \includegraphics[width=1.6cm, angle=-90]{delsanto06_fig3.ps}
      \caption{MOS1, MOS2, PN (obs. 401) and 2003 average ISGRI energy spectra 
       fitted with double component black body plus power law; 
the total model (continuous line), model components (dashed line) and residuals (bottom) are shown.}
         \label{fig:spec_bb_xmm}
   \end{figure}
%

\subsection{Hard X-ray temporal behaviour and broad-band spectral evolution}
The \1e temporal behaviour in the energy range 20--40 keV over 2 years of IBIS/ISGRI observations
is shown in Fig. \ref{fig:light}. 
It shows a rather constant flux (variation of about 20\%) on few months time scales, 
in agreement with results reported by Belanger et al. (2005).  
Because of its almost steady high energy behaviour, we fitted the \1e average IBIS/ISGRI spectrum collected in 2003
with the two non-simultaneous PN, MOS1 and MOS2 spectra of September 2000.

Since the 2--10 keV spectrum is well described
by an absorbed black body, we tried this model
also for the broad-band spectrum, with unacceptable results.
Indeed the high energy part of the spectrum by ISGRI required
the addition of an hard component modelised with a simple power law (see Fig. \ref{fig:spec_bb_xmm}
and Tab.~\ref{tab:bb_fit}).

We fitted the ACIS-I spectrum together with the 2004 ISGRI average
spectrum, and the XMM-Newton spectrum with the average
2003 ISGRI one. From the spectral parameters in Tab.~\ref{tab:bb_fit} it
seems to be an hardening of the source spectrum in 2004, compared to 2003. 
On the other hand, this
is only a weak indication, since fixing the column density of the 2004 spectrum
in the range 18--19$\times$$^{22}$~cm$^{-2}$ (as resulted in the 2003 broad band spectroscopy),
the fit is equally good, the photon index power-law steepens to 2.6--3.0, 
while the blackbody temperature remains hot at 2.3$\pm{0.2}$~keV. 

\begin{table*}[!ht]
\caption[]{Model parameters obtained by the broad-band spectral fit using 
\xmm~ (two observations) and ISGRI (2003 mean spectrum) and \chandra~ with the ISGRI spectrum averaged on 2004.
The power-law photon index $\Gamma$, the black-body temperature  $kT_{bb}$ and 
the column density $N_{H}$ are shown.
}
\begin{center}
\begin{tabular}[c]{lcccccc}
\hline\noalign{\smallskip} 
Observations  &  $N{\rm _H}$                  &$\Gamma$  & $kT_{bb}$ &  $F^{a}_{(2-10)}$   & $F^{b}_{(1-100)}$ & $\chi^2$/dof \\

       &($10^{22}$~cm$^{-2}$)  &                                     &  (keV) &                &               & \\
\noalign{\smallskip\hrule\smallskip}
\xmm(401) + IBIS(2003)              & $19.5 ^{+1.1}_{-0.9}$  & $3.1 \pm 0.1 $& $1.8 \pm 0.1 $ &  3.9    & 7.3$\pm 1.5$ & 468/375  \\ 
\noalign{\smallskip\hrule\smallskip}
\xmm(501) + IBIS(2003)               & $18.0^{+1.3}_{-1.0}$  & $3.3 \pm 0.1$ &  $1.6 \pm 0.1$ & 3.2 & 5.9$\pm 1.5$  & 392/323  \\
\noalign{\smallskip\hrule\smallskip}
\chandra(4500) + IBIS(2004)     & $12.7^{+2.9}_{-1.7}$  & $2.3^{+0.3}_{-0.2}$ &  $2.3^{+0.1}_{-0.3}$ & 3.3 & 6.6$\pm 1.5$  & 185/202  \\
\noalign{\smallskip\hrule\smallskip}
\end{tabular}
\small
\begin{itemize}
\small
\item[$^a$]The 2--10 keV flux of the unabsorbed best-fit model in units of 10$^{-10}$~erg~cm$^{-2}$~s$^{-1}$.
\small
\item[$^b$]The broad-band flux (1--100 keV) of the unabsorbed best-fit model in units of 10$^{-10}$~erg~cm$^{-2}$~s$^{-1}$.
\end{itemize}
\label{tab:bb_fit}
\end{center}
\end{table*}


\subsection{Luminosities}

In order to draw some conclusions on the source nature,
we assumed three possible distances for \1e, namely 8.5 kpc, 12 kpc and 20 kpc,
and we estimated the related luminosity (see Tab. \ref{tab:lumi}).
The black body plus power law model has been used for the 
unabsorbed flux evaluation. In particular we used the minimum unabsorbed flux we found,
i. e. 5.9$\times 10^{-10}$ erg cm$^{-2}$ s$^{-1}$ in the 1--100 keV energy band. 

Considering a 1.4 M$_{\odot}$ Neutron Star (NS) and a 10 M$_{\odot}$ Black Hole (BH),
the luminosity fractions in Eddington luminosity are also shown in the same table. 
\begin{table}[!t]
\caption[]{Luminosities of 1E 1743.1--2843 at different distances calculated both for a NS and BH.}
\begin{center}
\begin{small}
\begin{tabular}[c]{lccc}
\hline\noalign{\smallskip} 
Distance & L$_{1-100 keV}$ &  L/L$_{Edd}$ (BH) & L/L$_{Edd}$ (NS)\\
 & (erg/s) & & \\
\noalign{\smallskip\hrule\smallskip}
8.5 kpc & 5.2 $\times$ 10$^{36}$ & 0.3\% & 3\%\\ 
\noalign{\smallskip\hrule\smallskip}
12 kpc  & 1.0 $\times$ 10$^{37}$ & 0.8\% & 5\%\\
\noalign{\smallskip\hrule\smallskip}
20 kpc     &  2.9 $\times$ 10$^{37}$ & 2.0\%& 11\% \\
\noalign{\smallskip\hrule\smallskip}
\end{tabular}
\end{small}
\small
\label{tab:lumi}
\end{center}
\end{table}

\section{Discussion}
In this work, we have reported the first spectral analysis at high energy (above 20 keV)
of \1e, thanks to the high sensitivity and imaging capabilities of the \integral~imager IBIS.
The source is faint at high energy, i. e. roughly $1.7 \times 10^{-11}$ erg cm$^{-2}$ s$^{-1}$
(20--100 keV).
Our first broad-band spectral studies, during 2003 and 2004,
provided important constraints on the black body temperature and on the steepness 
of the power law component.
The latter may be interpreted as a thermal Comptonisation, 
although the low statistics at high energy hampered the use 
of more complex Comptonisation models. 
In the following we discuss the possible nature of the compact object.

\subsection{X-ray binary system scenario}
The lack of either pulsations or eclipses by previous observations 
and our monitoring, combined with the steep  
high energy component, could argue against a NS HMXB scenario
(as also suggested by Cremonesi et al. (1999)). 
Based on the long X-ray burster monitoring of \sax-WFC, in't Zand et al. (2004) reported on
the burst rate as a function of the persistent luminosity.
The trend in all bursting sources were consistent with a universal 
behaviour where the burst rate increases from the lowest luminosities to roughly 
1.3$\times$10$^{37}$ erg s$^{-1}$ and decreases above 2.9$\times$10$^{37}$ erg s$^{-1}$. 
At GC distances \1e would show the typical luminosity of bursters (see Tab. \ref{tab:lumi}),
hence the lack of type-I X-ray bursts during more than 20 years since the discovery is noteworthy.
Although the limited temporal coverage of the WFC (and also of the previous instruments), 
any type-I X-ray bursts should have been detected from a source with a typical bursting rate.
On the other hand, the lack of type-I burst detection is expected in the case that
\1e is a rare-burster (\cite{zand04} and reference therein). 
If so, it should lie behind the Galactic Centre,
at a distance providing L/L$_{Edd}$ of about 10\% .
Finally, the system could be a no-bursting source, a NS bright LMXB showing typical
luminosity of few $\sim 10^{37}$ erg/s, located behind the GC region, at a
distances larger than 15 kpc.

Our broad band spectral fitting requires a
a black body component at high temperatures 
and a faint and steep hard X-ray emission. 
In the BH scenario, these characteristics are typical of 
the canonical high/soft state (\cite{zdz04}, \cite{mcclint03}).
BH binaries in this state have been observed with luminosities 
not lower than 1-2\% L$_{Edd}$ (\cite{maccarone03}).
From this result and from the fact that luminosity is 1\% L$_{Edd}$ at 15 kpc,
we conclude the latter to be the lower limit for the distance. 

Up to now, nearly all LMXBs with persistent X-ray emission contain a NS (\cite{klis04}).
On the other hand, persistent emission from BH binaries is reported for sources in HMXB systems
(i. e., Cyg X--1, LMC X--1 and LMC X--3).
Moreover, LMC X--1 and LMC X--3 have been observed being usually in the high/soft state (\cite{haardt01})
and LMC X-1 showing a long term steady flux behaviour (\cite{mcclint03}).
In spite of such a similar behaviour, the disc temperature of \1e has been observed to be higher
than LMC X--1 and LMC X--3 (lower than 1 keV).
A recent work reports on systematic difference in the 
temperature of thermal components of NS vs BH binaries (\cite{remillard06}).
It has been found by these authors that temperatures for BHC are distributed around 1 keV, 
clearly disjoint from the $\sim$ 2 keV temperatures of NS. 
The hotter characteristic temperatures of NS systems is likely 
to originate in the boundary layer that forms when
accreting matter reaches the NS surface.
This statement could be a hint for the NS nature of \1e, 
even though a firm conclusion can not be established.

\subsection{Extra-galactic scenarios}
The AGN nature was also considered by Cremonesi et al. (1999), but now it seems
very unlikely, based on the spectral behaviour, which is quite different from 
this class of extra-galactic objects (\cite{nandra94}).
The possibility that the source is an Ultra Luminous X-rays source (ULX) can also be considered.
For a typical ULX luminosity of $10^{40}$~erg~s$^{-1}$ (see e.g. Colbert \& Ptak, 2002)
\1e should be located in a background galaxy at a distance of about
400~kpc. If ULXs are intermediate mass black holes (\cite{miller04}) with a mass around 100 solar masses, 
the accretion disk inner radius should be r$_{in}$$\times(cos(i)^{0.5})$ $\sim$1000~km, to be compared
with our spectral fitting results. Adopting a disc black body 
for the soft X-ray emission, we obtained a 
normalization in the range 0.2-0.5, which translates into
r$_{in}$$\times(cos(i)^{0.5})$ $\sim$ 20-30~km (assuming 400~kpc).
Thus, a ULX nature for \1e seems to be unlikely.

\begin{acknowledgements}
This work has been supported by the Italian Space Agency grant I/R/046/04. 
We thank Silvano Molendi for his help with \chandra~data analysis.  
The IBIS team at IASF-Rome thank Memmo Federici for its continuous support to the \integral~data archive.
MDS thanks Julien Malzac for precious BH physics discussion
and Luigi Piro for useful discussion on AGN. 
\end{acknowledgements}

\end{document}